\documentclass[conference,9pt]{IEEEtran}

\usepackage{times}
\usepackage{amsmath}
\usepackage{amssymb}
\usepackage{algorithm}
\usepackage{algorithmic}
\usepackage{color, array, colortbl}
\usepackage{graphicx}
\usepackage{multirow}
\usepackage{comment}
\usepackage{framed,color}
\definecolor{shadecolor}{rgb}{0.9,0.9,0.9}

\newcommand{\COMMENTED}[1]{}

\newcommand{\counter}{\textsc{C}}

\newcommand{\ONCONF}{\textsc{OnConf}}
\newcommand{\ONBR}{\textsc{OnBR}}
\newcommand{\OPTOFF}{\textsc{Opt}}
\newcommand{\OFFBR}{\textsc{OffBR}}
\newcommand{\ONTH}{\textsc{OnTh}}
\newcommand{\OFFTH}{\textsc{OffTh}}
\newcommand{\STAT}{\textsc{OffStat}}

\newcommand{\ALG}{\textsc{Alg}}
\newcommand{\OPT}{\textsc{Opt}}
\newcommand{\Cost}{\text{Cost}}

\newcommand{\BW}{\omega}
\newcommand{\LAT}{\lambda}
\newcommand{\CAP}{\omega}
\newcommand{\Load}{\text{load}}

\newtheorem{theorem}{Theorem}[section]

\newtheorem{definition}[theorem]{Definition}

\usepackage{algorithm}
\usepackage{algorithmic}

\newcommand{\ignore}[1]{}

\begin{document}






\title{On the Benefit of Virtualization:\\Strategies for Flexible Server Allocation}

\author {
   Dushyant Arora, Anja Feldmann,
   Gregor Schaffrath, Stefan
   Schmid\\
{\small Deutsche Telekom Laboratories / TU Berlin, Germany}\\
\texttt{\{darora,anja,grsch,stefan\}@net.t-labs.tu-berlin.de}\\
}

\date{}

\maketitle

\sloppy


\begin{abstract}
Virtualization technology facilitates a dynamic, demand-driven
allocation and migration of servers. This paper studies how the
flexibility offered by network virtualization can be used to improve
Quality-of-Service parameters such as latency, while taking into
account allocation costs. A generic use case is considered where
both the overall demand issued for a certain service (for example,
an SAP application in the cloud, or a gaming application) as well as
the origins of the requests change over time (e.g., due to time zone
effects or due to user mobility), and we present online and optimal
offline strategies to compute the number and location of the servers
implementing this service. These algorithms also allow us to study
the fundamental benefits of dynamic resource allocation compared to
static systems. Our simulation results confirm our expectations that
the gain of flexible server allocation is particularly high in
scenarios with moderate dynamics.\end{abstract}

%

\section{Introduction}
\label{sec_intro}

Virtualization is an intriguing paradigm which loosens the ties
between services and physical infrastructure. The gained flexibility
promises faster innovations, enabling a more diverse Internet and
ensuring coexistence of heterogeneous virtual network (VNet)
architectures on top of the shared substrate. Moreover, the dynamic
and demand driven allocation of resources may yield a ``greener
Internet'' without sacrificing (or, in the presence of the
corresponding migration technology: with improved!)
quality-of-service (QoS) / quality-of-experience (QoE).


This paper studies flexible offline and online allocation strategies
that achieve a good tradeoff between efficient resource usage and
QoS. We attend to a rather general use case which describes a system
providing a certain service to a dynamic set of users. The users'
access pattern can change over time (e.g., due to time zone effects
or due to user mobility), and hence, in order to ensure a low
latency, servers may adaptively be migrated closer to the current
origins of the requests; moreover, in peak hours, it can be
worthwhile to allocate additional resources.

This use case captures many different scenarios. For instance,
imagine an SAP application in the cloud which is accessed by
different users going online and offline over time, resulting in a
temporal change of the demand characteristics. Or, consider a mobile
provider which offers a gaming application to a set of mobile users
updating their location over time, and where access latency is of
prime concern.

\subsection{Our Contributions}

This paper studies strategies for flexible server allocation and
migration in a generic use case capturing various scenarios, ranging
from business applications such as SAP services in the cloud, to
entertainment applications such as mobile gaming. We present
algorithms which ensure a low access latency by adapting the
resources over time, while taking into account the corresponding
costs (communication cost, service interruption cost, allocation
cost, migration cost, and cost of running the servers). These
algorithms come in two flavors, exploring the two extremal
perspectives: online algorithms where decisions are done without any
information of future requests, and offline algorithms where the
(e.g.,  periodic) demand is known ahead of time. While optimal
solutions are computationally hard, there exist efficient adaptions,
which are discussed as well.

Our algorithms allow us to quantify the cost-benefit tradeoffs of
dynamic resource allocation, and shed light on fundamental questions
such as the use of migration compared to solutions with static
resources. For example, our simulations show that the overall cost
can be much higher if resources are static, in particular if the
demand dynamics is moderate.

%

\section{Model}\label{sec:model}

Our work is motivated by the recent advances in the field of network
virtualization. We are about to develop a prototype architecture,
and refer the reader to~\cite{visa09virtu} for more information on
this project. This section first provides some basic background and
subsequently identifies the major cost factors in our use case.

\subsection{Architecture and Use Case}

The virtualization architecture proposed in~\cite{visa09virtu}
distinguishes the following roles: The \emph{(Physical)
Infrastructure Provider (PIP)}, which owns and manages an
underlaying physical infrastructure (called ``substrate''); the
\emph{Virtual Network Provider (VNP)}, which provides bit-pipes and
end-to-end connectivity to end-users; and the \emph{Service Provider
(SP)}, which offers application, data and content services to
end-users.

As a generic use case, we consider a service provider offering a
service to mobile users which can benefit from the flexibility of
network and service virtualization. The goal of the service provider
is to minimize the round-trip-time of its service users to the
servers, by triggering migrations depending, e.g., on (latency)
measurements. Concretely, VNP and/or PIPs will react on the SP-side
changes of the requirements on the paths between server and access
points, and re-embed the servers accordingly.

\subsection{Graph Model and Access Cost}

Formally, we consider a substrate network $G=(V,E)$ managed by a
substrate provider (PIP), where $v\in V$ are the substrate nodes and
$e=(u,v)\in E$, with $u,v\in V$, are the substrate links; we will
refer to the total number of substrate nodes by $n=|V|$. Each
substrate node $v$ has a certain strength $\CAP(v)$ associated with
it (number of CPU cores, memory size, bus speed, etc.). A link is
characterized by a bandwidth capacity $\BW(e)$ and a latency
$\LAT(e)$. In addition to the substrate network, there is a set $T$
of external terminals (the mobile thin clients or the users) that
access $G$ by issuing requests for a given virtualized service
hosted on a set of virtual servers $S$ on $G$. We will assume that a
service is offered redundantly by up to $k=|S|$ servers.

In order for the clients in $T$ to access the servers $S$, a fixed
subset of nodes $A\subseteq V$ serve as \emph{Access Points} where
clients in $T$ can connect to $G$. Due to the request dynamics, the
popularity of access points can change frequently, which may trigger
the migration algorithm. We define $\sigma_t$ to be the multi-set of
requests at time $t$ where each element is a tuple $(a\in A, S\in
\mathcal{S})$ specifying the access point and the requested service
$S$. (For ease of notation, when clear from the context, we will
sometimes simply write $v\in \sigma_t$ to denote the multi-set of
access points used by the different requests.) Our main objective is
to shed light onto the trade-off between the access costs
$\Cost_{\text{acc}}$ of the mobile users to the service (delay of
requests), the server migration cost $\Cost_{\text{mig}}$, and the
cost $\Cost_{\text{run}}$ for running the servers: While moving the
servers closer to the requester may reduce the access costs and
hence improve the quality of service, it also entails the overhead
of migration; moreover, the more active servers, the more resources
needed (processing requests, CPU, storage, etc.).

In this paper, a simplified model is considered where the cost of
accessing the server, $\Cost_{\text{acc}}$, is given by the request
latency, i.e., the sum of the requests' latencies to the
corresponding servers (e.g., along the shortest paths on the
substrate network), plus the latency due to the server's $\Load$,
which, for server $v$ and at time $t$, is given by
$\Load(v,t)=f(\CAP(v),\eta(v,t))$, a function of the node strength
$\CAP(v)$ and the number of requests arriving at the servers hosted
by $v$ at time $t$, $\eta(v,t)$. For example, a simple model where
the load increases linearly would be $\Load(v,t)=\eta(v,t)/\CAP(v)$:
$$
\Cost_{\text{acc}}(t) = \sum_{r_t\in \sigma_t}
\text{delay}(r_t)+\sum_{v\in V}\Load(v,t).
$$
We will assume that requests are routed to the server of minimal
access costs.

\subsection{Server Model and Migration}

Each of the (at most $k$) servers can assume three different states:
\emph{not in use}, \emph{inactive}, and \emph{active}. If a server
is not in use, there are no costs. An inactive server comes at a
certain cost $R_i$ per time: this cost includes storing the
application software (e.g., the game) plus certain maintenance
costs. The running costs of an active server $R_a$ are larger, as
they also include CPU costs, maintaining state in the RAM, or
bandwidth costs. In order to startup a server which is not in use, a
fixed creation cost $c$ is assumed. For instance, this cost captures
the installation of the Linux box and the template (copy if already
on disk or download from an NFS share), configuration of the
template (e.g., setting up IP addresses manually or via DHCP),
starting the server etc. Finally, we assume that the cost of
changing from inactive to active state is negligible.

Also the cost of migration depends on many different factors. While
the operating systems is typically subject to replication (copy
Linux box from disk), the virtual server's configuration and
data/state component must be transmitted over the network. Besides
the cost for the bulk data transfer of the server state, there are
opportunistic costs that depend on whether the system supports live
migration or not, possibly some requests need to be routed to other
servers during migration, there can be periods of service
interruption, etc. We will consider a simplified scenario where
migration cost is described by a constant $\beta$ and where inactive
servers are not migrated.

How does $\beta$ relate to $c$?  It again depends on the scenario.
For example, $c\gg \beta$ for systems which support live migration
(almost no opportunistic or outage costs during migration), where
there is an NFS share and only the server state is migrated, and
where new servers need to be configured manually. On the other hand,
for example $c\ll \beta$ in systems where configuration is simple
and where migration happens over multiple provider domains. For the
formal description and analysis of the algorithms we will focus on
the more interesting case that $\beta<c$: if $\beta\geq c$,
migration is never beneficial, and the problem boils down on when
and where to create and delete servers; our algorithms could be
easily adapted for these situations as well.

Also note that in our model, migrating a server from node $v$ to an
empty node $v'$ costs $\beta$ and that subsequently, node $v$ is
empty. It is not possible to maintain a (for example inactive) copy
of the server at $v$ ``for free''; rather, this would require to set
up a new server which costs $c$, as the template needs
reconfiguration (e.g., new unique network addresses are needed).

However, let us emphasize that our approach is general enough to be
adopted in many alternative scenarios where the cost models are
slightly different, e.g., where server copies can be kept during
migration and $c$ denotes the cost of investing into an additional
server.

In order to clarify our model, we give three examples.\\
\noindent \emph{Example 1:} Assume we have three active servers
located at nodes $v_1$, $v_2$, $v_3$. When adding an additional
server at some node $v_4$, either: (1) if there is no inactive
server, an additional server has to be created at $v_4$ which costs
$c$ (we assume $\beta<c$); (2) if node $v_4$ already hosts an
inactive server, this server is simply activated and the cost is
zero; (3) otherwise, if there is an inactive server at some other
location $v_5$, this inactive server is migrated to $v_4$ which
costs $\beta$ (note that there will be no server at $v_5$ anymore).\\
\noindent \emph{Example 2:} Assume we have three servers hosted by
nodes $v_1$, $v_2$, $v_3$. In order to change to a configuration
where three servers are located at nodes $v_1$, $v_2$, and $v_4$,
either (1) if node $v_4$ is an inactive node, the cost is zero; (2)
if an inactive server at $v_5$ is migrated to $v_4$, the cost is
$\beta$, and the server from $v_3$ can become inactive (in our
algorithm, it will be added to a cache of inactive servers)---there
will be no server at $v_5$ anymore; (3) if the active server at
$v_3$ is migrated to $v_4$, the cost is $\beta$ as well and there is
no server at $v_3$ anymore.\\
\noindent \emph{Example 3:} Assume we have three servers hosted by
nodes $v_1$, $v_2$, $v_3$. When removing one server, i.e., when
changing to a configuration with two servers at nodes $v_1$ and
$v_3$, we do not incur any costs and the server at $v_2$ becomes
inactive (i.e., is added to the cache of inactive servers in our
algorithms).

\subsection{Request Model}

We next discuss the model for the terminal dynamics (e.g., due to
user mobility). One approach could be to assume arbitrary request
sets $\sigma_t$, where $\sigma_t$ is completely independent of
$\sigma_{t-1}$. However, for certain applications it may be more
realistic to assume that the requests move ``slowly'' between the
access points. We can distinguish between two different sources of
request dynamics: \emph{time zone effects} (users from different
countries access a service at different times of the day) and
\emph{user mobility}. Note that while users typically travel between
different cities or countries at a limited speed, these geographical
movements may not translate to the topology of the substrate
network. Thus, rather than modeling the users to travel along the
links of $G$, we may consider on/off models where a user appears at
some access point $a_1\in A$ at time $t$, remains there for a
certain period $\Delta t$, before moving to another arbitrary node
$a_2\in A$ at time $t+\Delta t$. Often, it is reasonable to assume
some form of correlation between the individual users' movement. For
example, in an urban area, workers commute downtown in the morning
and return to suburbs in the evening.

\subsection{Online and Offline Algorithms}

Typically, resources need to be allocated dynamically (i.e.,
\emph{online}) in virtual networks, without knowledge on the demand
or request sequence in advance. In order to focus on the main
properties and trade-offs involved in the the dynamic allocation and
migration problem, we assume a simplified online framework (see
also~\cite{visa10}). We assume a synchronized setting where time
proceeds in time slots (or \emph{rounds}). In each round $t$, a set
of $\sigma_t$ terminal requests arrive in a worst-case and online
fashion at an arbitrary set of access nodes $A$. Thus the embedding
problem is equivalent to the following synchronous game, where an
online algorithm $\ALG$ has to decide on the server allocation and
migration strategy in each round $t$, without knowing about the
future access requests. Concretely, in each round $t\geq 0$:
\begin{enumerate}

\item[1.] The requests $\sigma_t$ arrive at some access nodes $A$.

\item[2.] The online algorithm $\ALG$ pays the requests' access costs $\Cost_{\text{acc}}(t)$ to the corresponding servers.

\item[3.] The online algorithm $\ALG$ decides where in $G$ to allocate new or remove existing servers, which servers
should be active and which inactive, and where to migrate the
servers $S$. Accordingly it incurs running costs
$\Cost_{\text{run}}(t)$ as well as migration costs
$\Cost_{\text{mig}}(t)$.
\end{enumerate}
Observe that as we assume that the requests during one time slot are
much cheaper than a migration operation, our results also apply for
a scenario where the last two steps are reordered (see
also~\cite{visa10}).

To evaluate the efficiency of an online algorithm, its performance
is often compared to the performance of a (sometimes hypothetical)
optimal offline algorithm for the given request sequence. The ratio
of the two costs is called the \emph{competitive ratio}.

\section{Online Algorithms}\label{sec:companalysis}

This section presents strategies to allocate and migrate resources
in an online fashion---without knowing the future request
pattern---depending on the observed origins of the requests and the
load. A natural idea is to pursue a server configuration approach
which gives a general class of online algorithms.
\begin{definition}[Configuration]\label{defn:configuration}
A \emph{configuration} $\gamma$ describes, for each server, whether
it is \emph{not in use}, \emph{inactive}, or \emph{active}. In case
of inactive and active servers, $\gamma$ specifies where---i.e., on
which node---the server is located.
\end{definition}

In some sense, the single server algorithm proposed in~\cite{visa10}
can be regarded as a special case of this idea. Generalizing the
algorithm of~\cite{visa10} gives the following algorithm $\ONCONF$:
\begin{shaded}
$\ONCONF$ uses a counter $\counter(\gamma)$ for each configuration
$\gamma$. Time is divided into \emph{epochs}. In each epoch
$\ONCONF$ monitors, for each configuration $\gamma$, the cost of
serving all requests from this epoch by servers kept in
configuration $\gamma$, including the access costs (latency plus
induced load) of the requests, the server running costs, and
possible creation costs. $\ONCONF$ stores this cost in
$\counter(\gamma)$. The servers are kept in a given configuration
$\widehat{\gamma}$ until $\counter(\widehat{\gamma})$ reaches
$k\cdot c$. In this case, $\ONCONF$ changes to a configuration
$\widehat{\gamma}'$ chosen uniformly at random among configurations
with the property $\counter(\gamma) < k\cdot c$. If there is no such
configuration left, we do not migrate and the epoch ends in that
round; the next epoch starts in the next round and the counters
$\counter(\gamma)$ are reset to zero.
\end{shaded}

We can implement $\ONCONF$ in such a way that inactive servers are
managed by a queue of constant size. Inactive servers in the queue
are managed in a FIFO manner (older servers are replaced first); in
addition an inactive server expires after $x$ epochs for some
constant parameter $x$. (This also means that in configurations
$\gamma$ in $\ONCONF$, inactive servers are not included.)

Observe that during an epoch, $\ONCONF$ goes through at most
$O(k\log n)$ many configuration changes (or epochs), as there are
$\sum_{i=1}^k \binom{n}{i}$ many configurations. The cost per epoch
is at most $k\cdot c$, and it is easy to see that---under an overly
pessimistic perspective---also an optimal offline algorithm has cost
at least $\beta$ per $k$ epochs, so in competitive analysis
parlor~\cite{visa10}, we have a competitive ratio of at most
$O(c/\beta\cdot k^3\log{n})$.

Clearly, $\ONCONF$ needs to be optimized in many respects. For
instance, it can make sense to switch between ``close'' (with
respect to costs) configurations only, or to deterministically
switch to the configuration with the lowest counter. However, the
main problem of $\ONCONF$ is different: due to the configuration
complexity, the runtime is only acceptable for a small number of
servers $k$. Therefore, rather than discussing possible
optimizations, we concentrate on efficient variants for $\ONCONF$.

\subsection{Efficient Online Algorithms}\label{sec:heuristics}

There are several ways to speed up $\ONCONF$ such as clustering
approaches where optimal configurations are only considered on a
cluster granularity, or sampling approaches where, e.g., only $k$
configurations are tracked, one for each possible number of current
servers. In the following, we will focus on a sequential
best-response variant of $\ONCONF$ called $\ONBR$: updating one
server after another greatly reduces the configuration complexity,
while flexibility is maintained.
\begin{shaded}
$\ONBR$ starts in an arbitrary configuration, e.g., hosting one
server at the network center. Time is divided into epochs, and an
epoch ends when the total cost accumulated during this epoch
(including access cost and running cost) reaches a threshold
$\theta$. Then, $\ONBR$ changes to the cheapest (w.r.t.~the passed
epoch and including access, migration, running, and creation cost)
configuration among the following: (1) $\gamma$ (no change), (2)
$\gamma$ but where one server $s$ is migrated to a different
location ($O(n)$ options), (3) $\gamma$ but where one server $s$
becomes inactive ($O(k)$ options), (4) $\gamma$ but where one
inactive server $s$ becomes active, or a new active server $s$ is
created  ($O(n)$ options). Inactive servers are organized in a queue
of constant size where the oldest server in the queue is the first
to be replaced---i.e., this server will no longer be in use (in our
simulations: size $3$). In case a new server is created at an empty
node, the oldest inactive server from the queue is migrated to the
corresponding node. Inactive servers in the queue expire after $x$
epochs for some parameter $x$ ($x=20$ in our simulation).
\end{shaded}

$\ONBR$ requires a good choice of the parameter $\theta$ in order to
trade-off flexibility and optimality. An alternative intuitive
approach is to add new servers when the access cost is higher than
the total running cost (w.r.t.~a certain threshold). This is
automatized by the following algorithm $\ONTH$.
\begin{shaded}
$\ONTH$ starts in an arbitrary configuration, e.g., hosting one
server at the network center. Time is divided into small and large
epochs: a small epoch ends when we have accumulated a cost of
$y\cdot \beta$ in a given configuration for some constant parameter
$y$ ($y=2$ in our simulations), and a large epoch ends when the
accumulated access cost is larger than the accumulated running cost
(of the active servers); concretely, we will use the following
condition: $\Cost_{\text{acc}}/(k_{\text{cur}}+1) -
\Cost_{\text{run}}> c$, where $k_{\text{cur}}$ denotes the current
number of active servers. When a small epoch ends $\ONTH$ changes to
the cheapest (w.r.t.~the passed epoch and including access,
migration, and running cost) configuration among the following: (1)
$\gamma$ (no change), (2) $\gamma$ but where one server $s$ is
migrated to a different location (the server at the migration origin
becomes inactive), (3) $\gamma$ but where one server $s$ becomes
inactive (at most $k$ options). Inactive servers are organized in a
first-in-first-out (FIFO) queue of constant size (in our
simulations: size $3$), i.e., inactive servers which fall out of the
queue are no longer in use. Inactive servers in the queue expire
after $y\cdot\beta$ rounds for some parameter $x$ ($x=20$ in our
simulation). When a large epoch ends, a new server is activated at
an optimal position with respect to the access cost of the latest
large epoch.
\end{shaded}

Note that both $\ONBR$ and $\ONTH$ have the appealing property that
in case of constant demand, they will eventually converge to a
stable configuration.

\section{Offline Algorithms}\label{sec:optoff}

While in the worst case, the decisions when and where to migrate
servers typically needs to be done \emph{online}, i.e., without the
knowledge of future requests, there can be situations where it is
interesting to study which migration pattern would have been good
\emph{at hindsight}. For example, if it is known that the requests
follow a regular pattern (e.g., a periodic pattern per day or week),
it can make sense to compute the migration strategy offline and
apply it in the future. While Section~\ref{sec:companalysis} assumed
an extreme standpoint and only discussed algorithms that do not have
any knowledge of future requests, in the following, the other
extreme standpoint is explored where future demand is completely
known. Please note that there is also another reason why designing
offline algorithms explicitly may be of interest, namely to compute
competitive ratios (see~\cite{visa10}) in simulations.

\subsection{Optimal Offline Algorithm}

This section presents an optimal offline algorithm $\OPTOFF$ for our
resource allocation optimization problem. It turns out that offline
strategies can be computed for many different scenarios, and we
describe a very general algorithm here.

Algorithm $\OPTOFF$ is based on dynamic programming techniques and
also uses the concept of configurations (cf
Definition~\ref{defn:configuration}). Recall that given a
configuration $\gamma$, access costs $\Cost_{\text{acc}}$, migration
costs $\Cost_{\text{mig}}$, and the running costs
$\Cost_{\text{run}}$ over time can be computed.

$\OPTOFF$ exploits the fact that the migration problem exhibits an
optimal substructure property: Given that at time $t$, the $k$
servers are in a configuration $\gamma$, then the most
cost-efficient \emph{path} (migrations, activation and deactivation
of servers, creation, etc.) that leads to this configuration
consists solely of optimal sub-paths. That is, if a cost minimizing
path to configuration $\gamma$ at time $t$ leads over a
configuration $\gamma'$ at time $t'<t$, then there cannot be a
cheaper migration sub-path that leads to $\gamma'$ at time $t'$ than
the corresponding sub-path.

$\OPTOFF$ essentially fills out a matrix
$opt[\text{time}][\text{configuration}]$ where $opt[t][\gamma]$
contains the cost of the minimal path that leads to a configuration
where the servers satisfy the requests of time $t$ in a
configuration $\gamma$. Recall from
Definition~\ref{defn:configuration} that a configuration $\gamma$
describes for each virtual server $s$ at which physical node $v$ it
is hosted and whether $s$ is \emph{not in use}, \emph{inactive}, or
\emph{active}.

Assume that in the beginning, the system is located in configuration
$\gamma_0$. Thus, initially, $opt[0][\gamma]=\Cost(\gamma_0
\rightarrow \gamma) + \Cost_{\text{run}}(\gamma) + \left[\sum_{v\in
\sigma_0}\Cost_{\text{acc}}(v,\gamma) \right]$, where
$\Cost(\gamma_1 \rightarrow \gamma_2)$ denotes the cost of changing
from configuration $\gamma_1$ to $\gamma_2$ (cost of migrations,
creation costs, etc.), $\Cost_{\text{run}}(\gamma)$ denotes the cost
of running the inactive and active servers for one time unit in
configuration $\gamma$, and $\sum_{v\in
\sigma_0}\Cost_{\text{acc}}(v,\gamma)$ denotes the access costs
(request latency and server load) resulting from the requests of
$\sigma_0$ accessing the active servers in configuration $\gamma$.
(W.l.o.g., we assume that the cost $\Cost_{\text{acc}}$ contains the
first wireless hop from terminal to substrate network.)

For $t>0$, we find the optimal values $opt[t][\gamma]$ by
considering the optimal paths to any configuration $\gamma'$ at time
$t-1$, and adding the migration cost from $\gamma'$ to $\gamma$.
That is, in order to find the optimal cost to arrive at a
configuration with servers at $\gamma$ at time $t$:
\begin{shaded}
\begin{footnotesize}
$$
\min_{\gamma'}\left[opt[t-1][\gamma']+\Cost(\gamma'\rightarrow
\gamma)+\Cost_{\text{run}}(\gamma)+\sum_{v\in
\sigma_t}\Cost_{\text{acc}}(v,\gamma)\right]
$$
\end{footnotesize}
\end{shaded}
\noindent where we assume that $\Cost_{\text{acc}}$ includes the
first (wireless) hop of the request from the terminal to the
substrate network, and where
$\Cost(\gamma\rightarrow\gamma)=0~~\forall \gamma$.

Note that $\OPTOFF$ is not an online algorithm, although $opt[t]$
does not depend on future requests: in order to reconstruct the
optimal migration strategy at hindsight, the configuration of
minimal cost after the last request is determined, and from there,
the optimal path is given by recursively finding the optimal
configuration at time $t-1$ which led to the optimal configuration
at time $t$.

\subsection{Efficient Offline Algorithms}

Unfortunately, the computational complexity of $\OPTOFF$ is rather
high for scenarios with many servers. We believe that it is
difficult to significantly reduce the runtime. Again, clustering or
sampling heuristics may be used to speed up the computations (which
may come at a loss of allocation quality).

There is an interesting and natural adaption of the best response
strategies of Section~\ref{sec:companalysis}: $\OFFBR$ is similar to
$\ONBR$, but rather than switching to the configuration of lowest
cost w.r.t.~the passed epoch, we switch to the configuration of
lowest cost in the \emph{upcoming epoch}! A similar transformation
can be done from $\ONTH$ to $\OFFTH$: we simply compute optimal
strategies of small epochs at hindsight. This yields an acceptable
runtime.

\section{Simulations}\label{sec:simulations}

We conducted several experiments and in the following, we report on
our simulation results and main insights.

\subsection{Set-up}

We conducted experiments on both artificial \emph{Erd\"os-R\'{e}nyi
graphs} random graphs (with connection probability 1\%) as well as
more realistic graphs taken from the \emph{Rocketfuel
project}\footnote{For maps and data, see
\texttt{http://www.cs.washington.edu}\\\texttt{/research/networking/rocketfuel/}.}~\cite{rocketfuel2,rocketfuel}
(including the corresponding latencies for the access cost). To
simulate $\OPT$, we constrain ourselves to line graphs.  If not
stated otherwise, we assume that link bandwidths are chosen at
random (either T1 (1.544 Mbit/s) or T2 (6.312 Mbit/s)), that
$\beta=40$ and that $c=400$; for experiments with $\beta>c$, we set
$\beta=400$ and $c=40$.

As real traffic patterns are confidential and cannot be published,
we consider two different simplified, artificial scenarios,
reflecting the two main reasons for request dynamics: \emph{time
zone effects} (users from different countries access the service at
different times of the day) and \emph{user mobility}.

\begin{shaded}
\textbf{Time Zones Scenario:} This scenario models an access pattern
that can result from global daytime effects. We divide a day into
$T$ time periods. For each time $t$, $p\%$ of all requests originate
from a node chosen uniformly at random from the substrate network
(we assume that these locations are the same each day). (Recall that
the substrate topology does not necessarily reflect the geographic
situation, but note that the uniform random choice is still
pessimistic.) The sojourn time of the requests at a given location
is constant and given by a parameter $\tau$. In addition, there is a
background traffic: The remaining requests originate from nodes
chosen uniformly at random from all access points.
\end{shaded}

\begin{shaded}
\textbf{Commuter Scenario:} This scenario models an access pattern
that can result from commuters traveling downtown for work in the
morning and returning back to the suburbs in the evening. The
scenario comes in two flavors: one with \emph{static demand} and one
with \emph{dynamic demand}.
\begin{enumerate}
\item \emph{Static Load}:
We use a parameter $T$ to model the \emph{frequency} of the changes.
At time $t$ mod $T < T/2$, there are $2^{t \text{ mod } T}$ requests
originating from access points chosen uniformly at random around the
center of the network. In the second half of the day, i.e., for $t
\in [T/2,...,T]$, the pattern is reversed. Then a new day starts.
The total number of requests per round is fixed to $2^{T/2}$. At
time $t_i < T/2$, the requests originate from $p = 2^{t_i \text{ mod
} T}$ of all access points including the network center ($2^{T/2}/p$
requests per access point), until single requests originate from
$2^{T/2}$ access points. Then, the same process is reversed until
all $2^{T/2}$ requests originate from a single access point: the
network center. We assume that the time period between $t_i$ and
$t_{i+1}$ is given by a fixed parameter $\lambda$.
\item \emph{Dynamic Load}: The total number of requests per round is not fixed to $2^{T/2}$.
At time $t_i < T/2$, the requests originate from $p = 2^{t_i \text{
mod } T}$ of all access points including the network center (one
request per access point), until single requests originate from
$2^{T/2}$ access points. Then, the same process is reversed until we
have a single request originating from a single access point: the
network center. We assume that the time period between $t_i$ and
$t_{i+1}$ is fixed and we denote it by parameter $\lambda$.
\end{enumerate}
\end{shaded}

\subsection{Experiments}

\begin{figure}[t]
\begin{center}
\includegraphics[width=0.83\columnwidth]{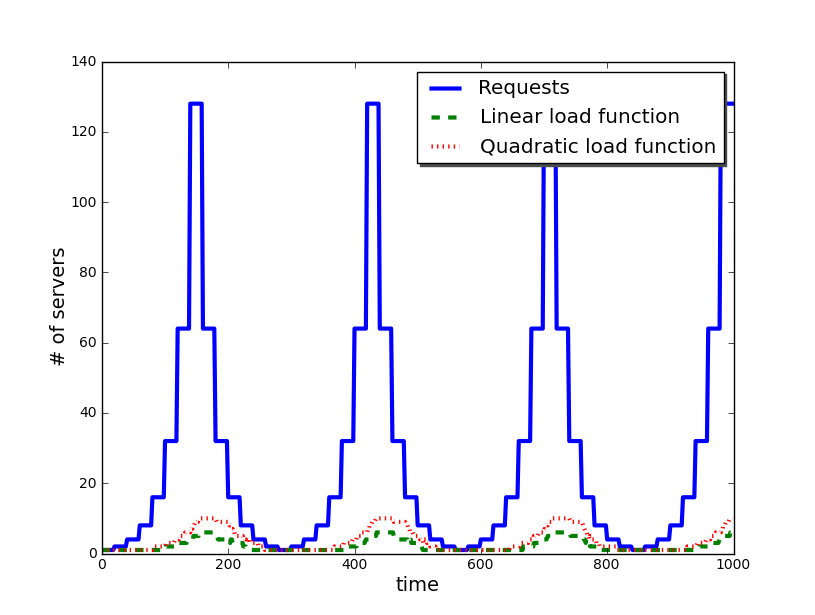}\\
\caption{Exemplary execution of $\ONTH$ in commuter scenario with
dynamic load. In this setting, the runtime was $1000$ rounds,
$T=14$, we considered a network of size $1000$, and set
$\lambda=20$.}\label{fig:commuter_scenario_dl_load}
\end{center}
\end{figure}

\begin{figure}[t]
\begin{center}
\includegraphics[width=0.83\columnwidth]{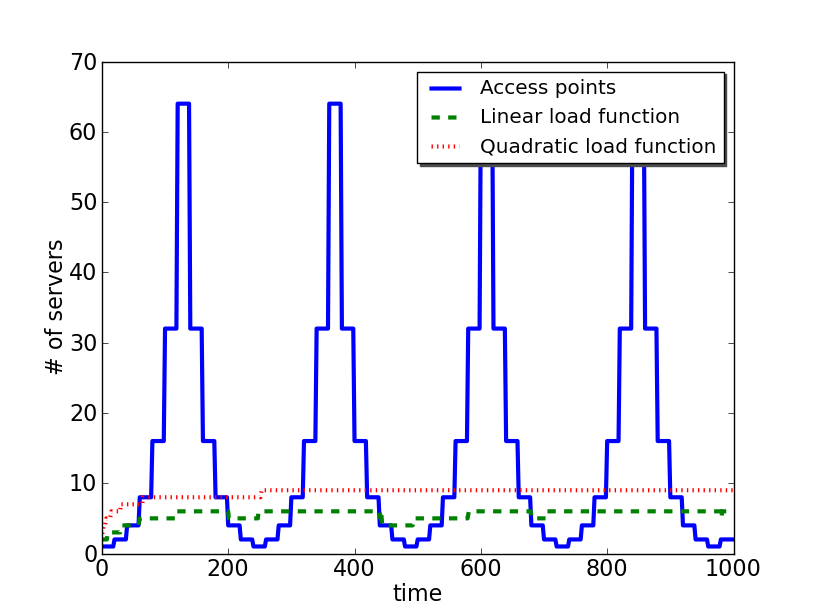}\\
\caption{Exemplary execution of $\ONTH$ in commuter scenario with
static load. In this setting, the runtime was $1000$ rounds, $T=12$,
we considered a network of size $500$, and set
$\lambda=20$.}\label{fig:commuter_scenario_sl_load}
\end{center}
\end{figure}

The main objective of our algorithms is to adapt to dynamically
changing demands in an efficient manner. Where to allocate or
migrate how many servers depends on the origins and the size of the
requests, and also on how access cost increases as a function of
load. As a motivation, consider the exemplary executions of $\ONTH$
depicted in Figure~\ref{fig:commuter_scenario_dl_load} for linear
and quadratic load functions. It can be seen intuitively that
$\ONTH$ reacts as desired in this example, reacting to higher loads
(either due to higher demand or steeper load functions) by
allocating more servers.
Figure~\ref{fig:commuter_scenario_sl_load} shows the same execution
for the static load variant. Initially, the system converges quickly
to a certain number of servers. It can also be seen that the number
of servers in this scenario is more or less independent of the
number of access points from which the given requests originate
from, and that a quadratic load model requires more servers.

In the following, after these motivating examples, the performance
of our algorithms is studied more systematically by considering the
dependencies of the performance on the various parameters. A first
set of experiments compares the performance of $\ONBR$ and $\ONTH$.
For $\ONBR$, a threshold $2\cdot c$ is considered. In addition, we
experimented with a variant where the threshold also depends on the
length $\ell$ of the preceding epoch, i.e., $2\cdot c/\ell$---in
some sense, a shorter epoch denotes faster changes in the requests,
and hence, the system should adapt more quickly. We will refer to
the two variants of $\ONBR$ by ``\emph{fixed}'' and ``\emph{dyn}''.
Clearly, many other variants are possible, e.g., where the threshold
depends on the variance (over time) of the access cost. However,
since $\ONTH$ typically outperformed $\ONBR$ and as $\ONTH$ requires
less parameter, we do not discuss the $\ONBR$ variants in more
detail but will focus on $\ONTH$. Moreover, note that for $\beta>c$,
migration is never useful, and the three algorithms coincide; in
this case, we will simply consider $\ONBR$ with fixed threshold $2c$
in this case.

\begin{figure}[t]
\begin{center}
\includegraphics[width=0.83\columnwidth]{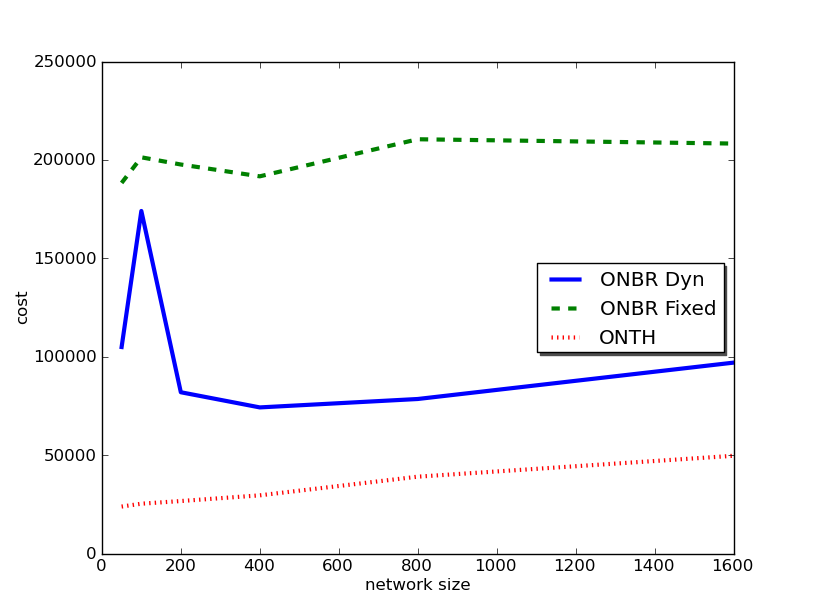}\\
\caption{Cost of different algorithms in commuter scenario with
dynamic load as a function of network size. The runtime was 500
rounds, $\lambda=10$, and we averaged over 5 runs. Note that $T$
increases with network size in our
model.}\label{fig:cost_commuter_scenario_dl}
\end{center}
\end{figure}

\begin{figure}[t]
\begin{center}
\includegraphics[width=0.83\columnwidth]{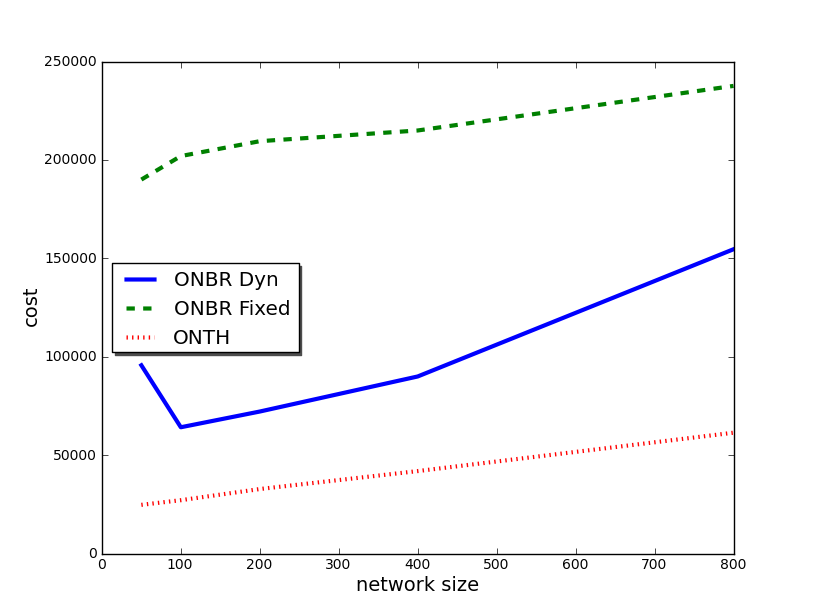}\\
\caption{Like Figure~\ref{fig:cost_commuter_scenario_dl}, but with
static load.}\label{fig:cost_commuter_scenario_sl}
\end{center}
\end{figure}

\begin{figure}[t]
\begin{center}
\includegraphics[width=0.83\columnwidth]{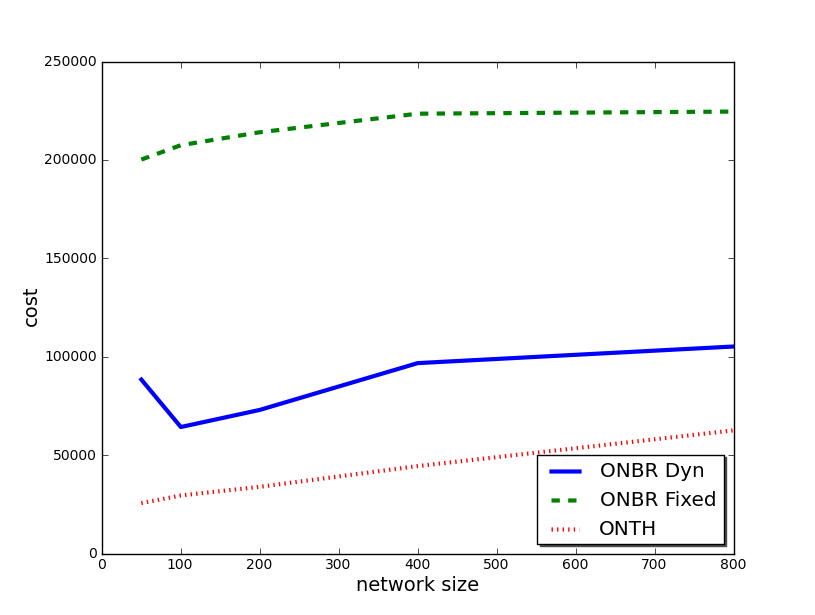}\\
\caption{Like Figure~\ref{fig:cost_commuter_scenario_dl}, but for
time zone scenario.}\label{fig:cost_time_zone}
\end{center}
\end{figure}

Figure~\ref{fig:cost_commuter_scenario_dl} compares the cost of the
different algorithms in the commuter scenario with dynamic load and
as a function of network size. We can see that $\ONTH$ has lower
costs than the $\ONBR$ variants, although the cost increase slightly
faster with the number of nodes in the substrate network.
Figure~\ref{fig:cost_commuter_scenario_sl} and
Figure~\ref{fig:cost_time_zone} show the same results for static
load commuter scenario and the time zone scenario, respectively.

\begin{figure}[t]
\begin{center}
\includegraphics[width=0.83\columnwidth]{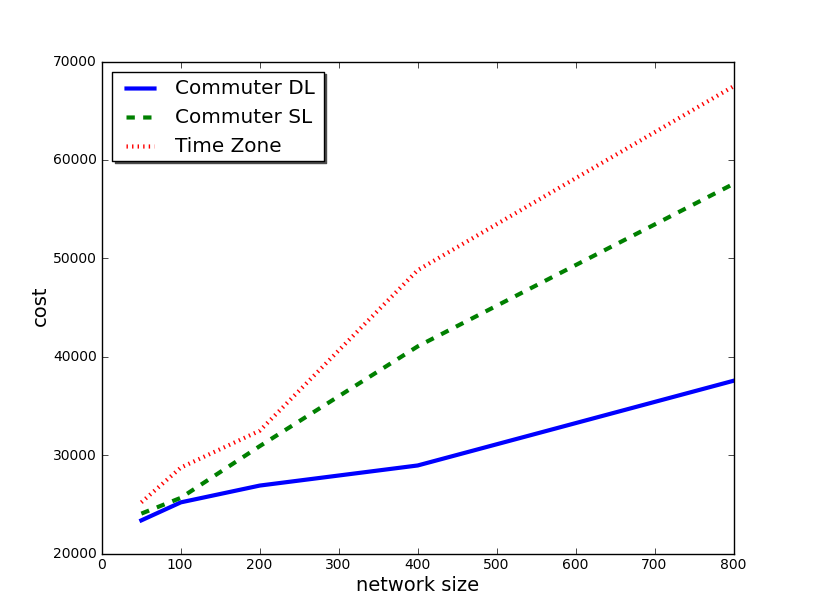}\\
\caption{Comparison of costs incurred by $\ONBR$ in different
scenarios as a function of network size (runtime 500 rounds,
$\lambda=10$, $\beta=400$, $c=40$, and averaged over 5
runs.)}\label{fig:all_three_scenario_cost_beta_gt_c}
\end{center}
\end{figure}

Of course, there are different costs involved in the different
scenarios. Figure~\ref{fig:all_three_scenario_cost_beta_gt_c} shows
how these costs relate to each other in the case of $\ONBR$ in a
scenario where $\beta>c$. (Recall that for $\beta>c$, the three
algorithms coincide, and we simply consider $\ONBR$ with fixed
threshold $2c$ in this case.)

\begin{figure}[t]
\begin{center}
\includegraphics[width=0.83\columnwidth]{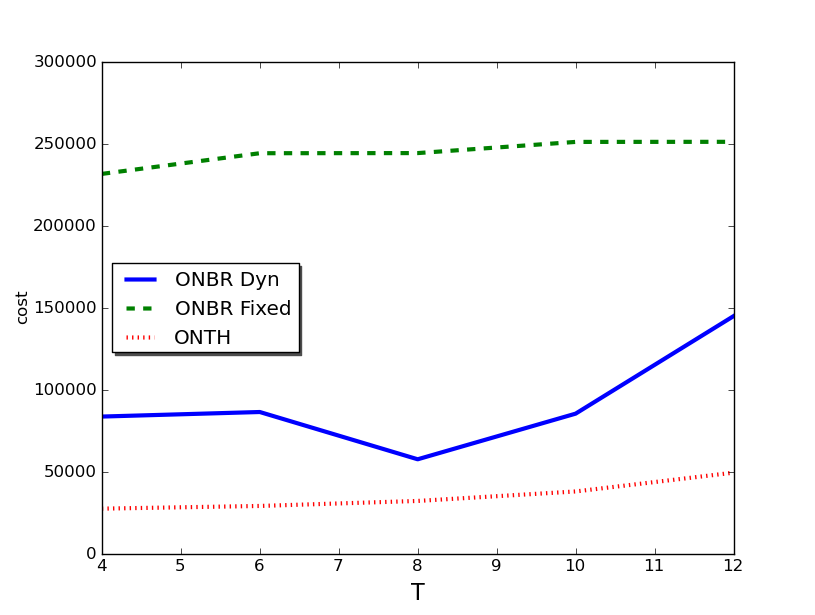}\\
\caption{Cost as a function of $T$ for different strategies in a
commuter scenario with static load (runtime 600, $\lambda=20$,
network size 1000, averaged over 10
runs.)}\label{fig:commuter_scenario_sl_vary_frequency}
\end{center}
\end{figure}

Figure~\ref{fig:commuter_scenario_sl_vary_frequency} studies the
cost of the different algorithms as a function of $T$ in the
commuter scenario with static load. Also here, $\ONTH$ always yields
the best performance. (Note that the cost slightly increases with
$T$, which is due to the fact that the request horizon is larger for
larger $T$ in our scenario.)

\begin{figure}[t]
\begin{center}
\includegraphics[width=0.83\columnwidth]{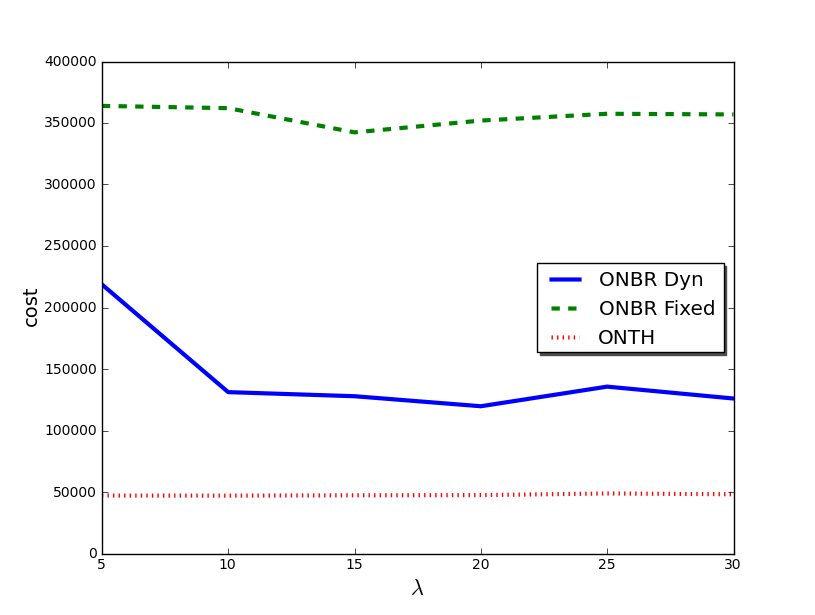}\\
\caption{Cost as a function of $\lambda$ in commuter scenario with
dynamic load (runtime 900 rounds, $T=10$, network size $200$,
averaged over $10$
runs).}\label{fig:commuter_scenario_dl_vary_stayduration}
\end{center}
\end{figure}
\begin{figure}[t]
\begin{center}
\includegraphics[width=0.83\columnwidth]{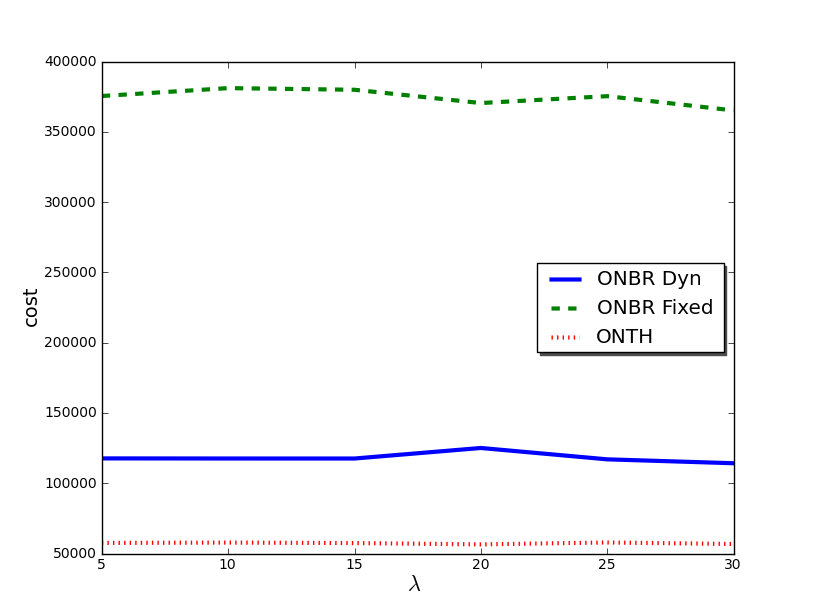}\\
\caption{Cost as a function of $\lambda$ in commuter scenario with
static load (runtime 900 rounds, $T=10$, network size 200, averaged
over 10 runs.)}\label{fig:commuter_scenario_sl_vary_stayduration}
\end{center}
\end{figure}
\begin{figure}[t]
\begin{center}
\includegraphics[width=0.83\columnwidth]{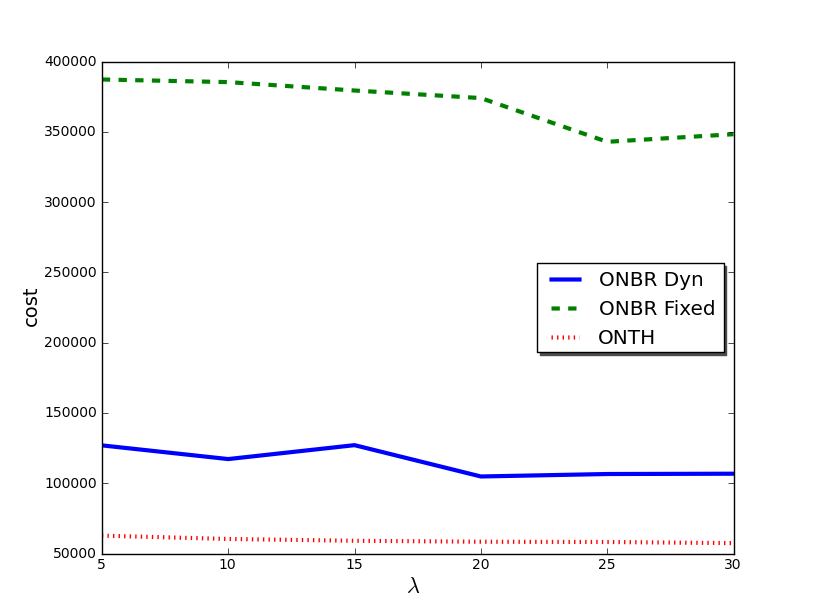}\\
\caption{Cost as a function of $\lambda$ in time zone scenario with
with $p=50\%$ (runtime 900 rounds, $T=10$, network size 200,
averaged over 10 runs).}\label{fig:time_zone_vary_stayduration}
\end{center}
\end{figure}

The total cost is more or less independent of $\lambda$, as shown in
Figure~\ref{fig:commuter_scenario_dl_vary_stayduration}, while
$\ONTH$ is better by a factor of approximately two.
Figure~\ref{fig:commuter_scenario_sl_vary_stayduration} presents the
same results for a static load scenario, and
Figure~\ref{fig:time_zone_vary_stayduration} presents the time zone
scenario. In the latter, the total cost decreases slightly with
$\lambda$, which is due to the fact that less migrations are needed
for larger $\lambda$.

\begin{figure}[t]
\begin{center}
\includegraphics[width=0.83\columnwidth]{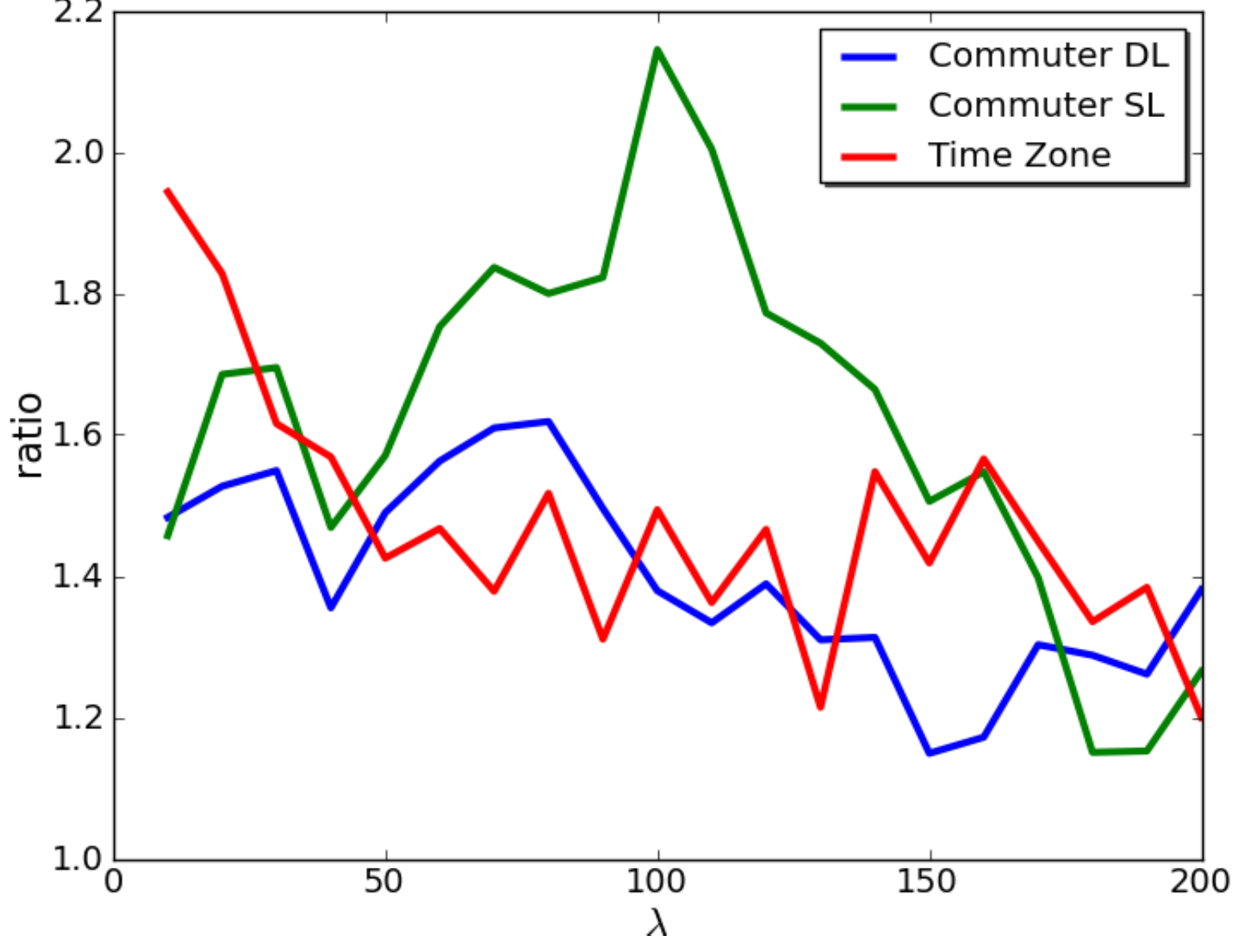}\\
\caption{Ratio of $\ONTH$ cost divided by $\OPT$ cost as a function
of $\lambda$, runtime 200 rounds, in a network with five nodes,
averaged over 10 runs.}\label{fig:compth}
\end{center}
\end{figure}

Our algorithms also allow us to get a
glimpse\footnote{Unfortunately, the networks for which we can run
these experiments are small.} onto the price of online decision
making, and more importantly, the question of when dynamic
allocation and migration technology is most useful. Regarding the
price of the lack of knowledge of future requests,
Figure~\ref{fig:compth} shows the competitive ratio of $\ONTH$ as a
function of $\lambda$. While the ratios are fairly low in all
scenarios, the static load commuter scenario reaches the highest
peak for an intermediate $\lambda$.

\begin{figure}[t]
\begin{center}
\includegraphics[width=0.83\columnwidth]{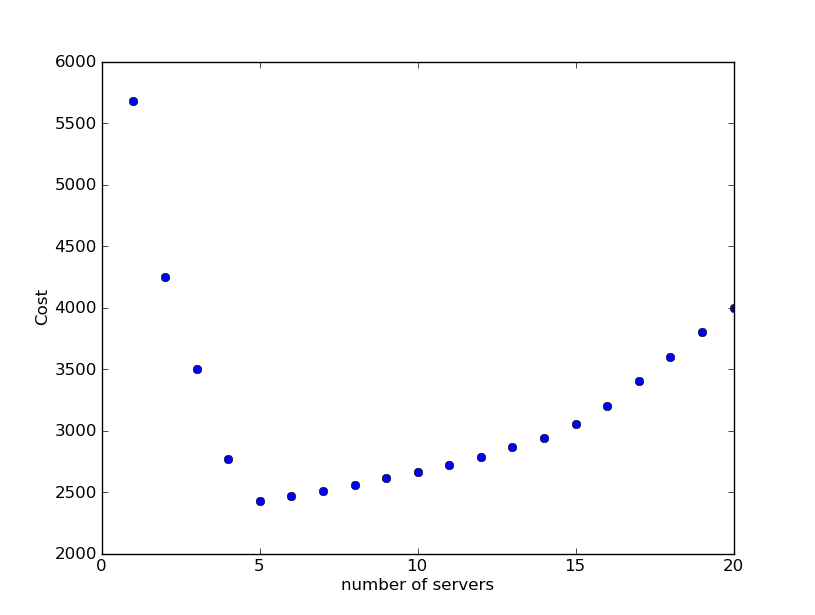}\\
\caption{$\STAT$ determines the best number of servers by minimizing
the total cost.}\label{fig:multiserver-scale-test}
\end{center}
\end{figure}

In order to shed light onto the benefits of migration, as a
reference point, we use the following static (but offline) algorithm
$\STAT$ which we will compare to an optimal offline algorithm with
migration.
\begin{shaded}
For a given request sequence $\sigma$, $\STAT$ determines the
optimal number of servers $k_{opt}$ as follows. For each
$i\in\{1,\dots,k\}$, we compute the cost of the following greedy
static configuration for $\sigma$: one active server
$j\in\{1,\dots,i\}$ after the other is placed greedily at the
location which yields the lowest cost for $\sigma$, given the
already placed servers $\{1,\dots,j-1\}$. $k_{opt}$ is defined as
the $i$ with minimal cost. Figure~\ref{fig:multiserver-scale-test}
illustrates how $\STAT$ computes the number of servers.
\end{shaded}

\begin{figure}[t]
\begin{center}
\includegraphics[width=0.83\columnwidth]{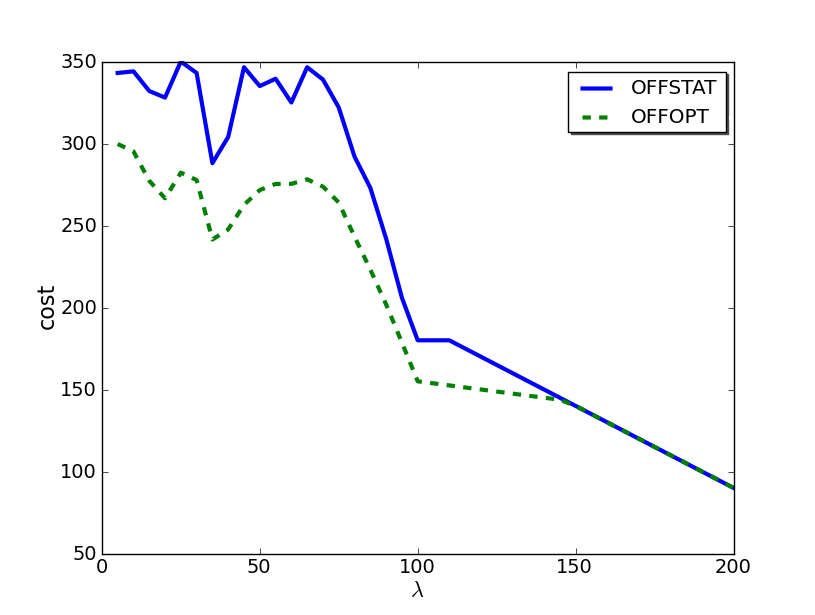}\\
\caption{The use of dynamic allocation in commuter scenario with
dynamic load as a function of $\lambda$. The experiment ran for 200
rounds in a network of five nodes where $T=4$, averaged over 10
runs.}\label{fig:opt_vs_stat_commuter_dl_vary_stayduration_lambda_equals_runtime_beta_gt_c}
\end{center}
\end{figure}

\begin{figure}[t]
\begin{center}
\includegraphics[width=0.83\columnwidth]{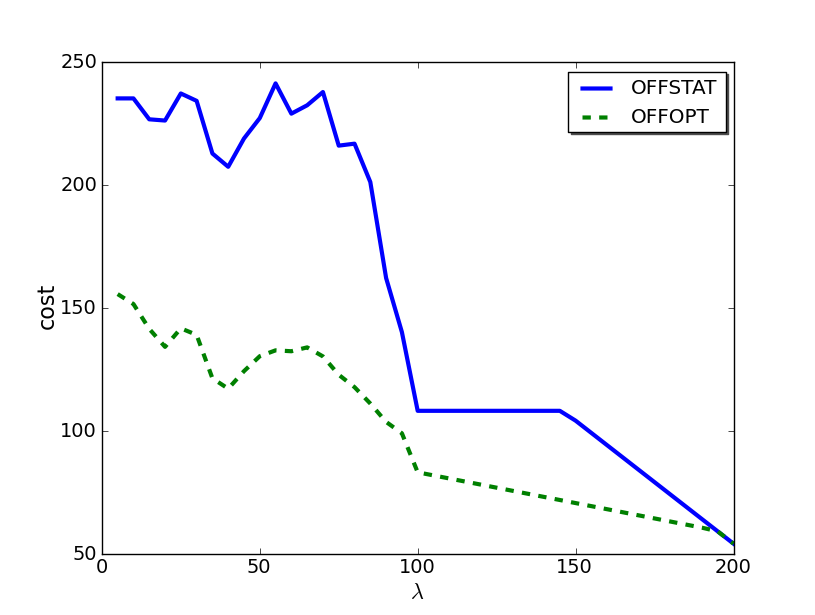}\\
\caption{Costs in dynamic load commuter scenario where $\beta=400$
and $c=40$ as a function of $\lambda$ (runtime 200 rounds, network
size five, $T=4$, averaged over 10
runs).}\label{fig:opt_by_stat_commuter_dl_vary_stayduration_lambda_equals_runtime}
\end{center}
\end{figure}

Figure~\ref{fig:opt_vs_stat_commuter_dl_vary_stayduration_lambda_equals_runtime_beta_gt_c}
shows the absolute costs of $\STAT$ and $\OPTOFF$ as a function of
$\lambda$ in the commuter scenario with dynamic load. As expected,
in less dynamic systems, the cost goes down, and the relative
advantage of the allocation and migration flexibility declines.
Figure~\ref{fig:opt_by_stat_commuter_dl_vary_stayduration_lambda_equals_runtime}
shows the same result for the $\beta>c$ scenario.

\begin{figure}[t]
\begin{center}
\includegraphics[width=0.83\columnwidth]{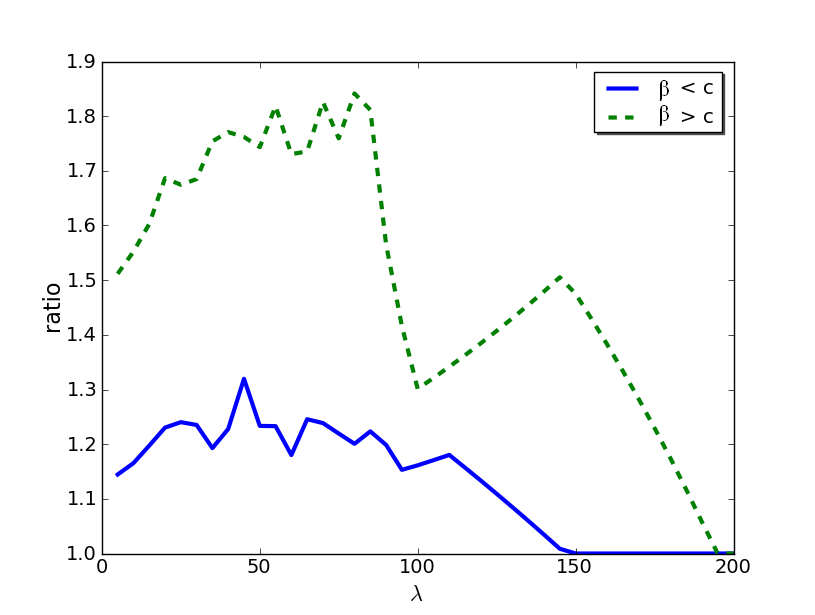}\\
\caption{Ratio of $\STAT$ and $\OPTOFF$ costs in dynamic load
commuter scenario as a function of $\lambda$, where runtime was 200
rounds, $T = 4$, network size 5, and averaged over 10
runs.}\label{fig:opt_vs_stat_commuter_dl_vary_stayduration_lambda_equals_runtime}
\end{center}
\end{figure}

In contrast to the absolute costs, the relative costs capture the
use of dynamic allocation more directly.
Figure~\ref{fig:opt_vs_stat_commuter_dl_vary_stayduration_lambda_equals_runtime}
plots the ratio of the total cost incurred by $\STAT$ by the total
cost incurred by $\OPT$ in the dynamic load commuter scenario as a
function of $\lambda$. As can be seen, for very high dynamics as
well as for very low dynamics, the flexibility of $\OPT$ is of
limited benefit. However, for moderate dynamics, it is worthwhile
for $\OPT$ to exploit the request patterns, and a better performance
can be achieved (up to a factor of two). This result also meets our
expectation. Interestingly, however, it turns out that $\OPT$ is
relatively better if $\beta>c$, i.e., in scenarios where migration
is never an option.

\begin{figure}[t]
\begin{center}
\includegraphics[width=0.83\columnwidth]{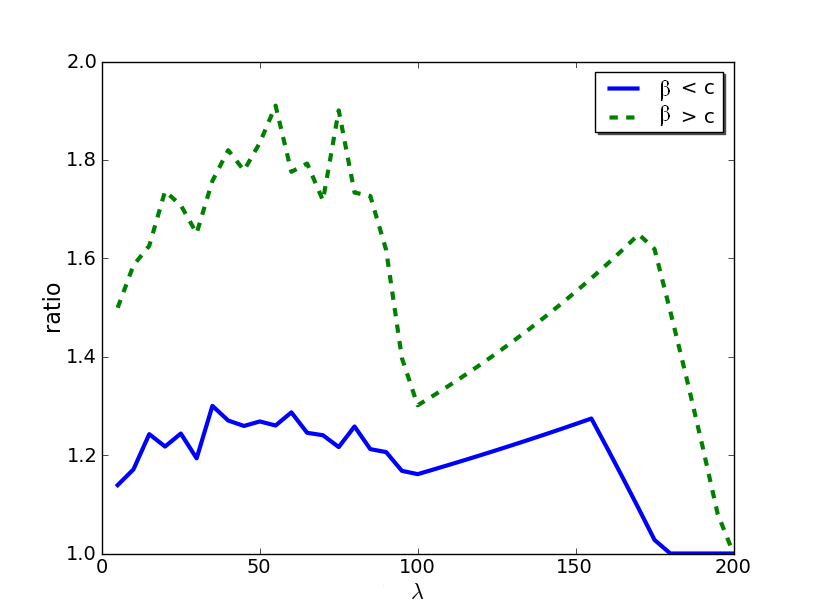}\\
\caption{Ratio of $\STAT$ and $\OPT$ costs in static load commuter
scenario as a function of $\lambda$, where runtime was 200 rounds,
$T=4$, network size five, and averaged over 10
runs.}\label{fig:opt_by_stat_time_zone_vary_stayduration_lambda_equals_runtime}
\end{center}
\end{figure}

Also in a commuter scenario with static load
(Figure~\ref{fig:opt_by_stat_time_zone_vary_stayduration_lambda_equals_runtime}),
$\beta<c$ yields the lower ratio, fluctuating more or less
constantly around 1.2 until it goes down to one for static access
patterns. For $\beta>c$, the ratio goes up to almost two for
intermediate $\lambda$ values.

\begin{figure}[t]
\begin{center}
\includegraphics[width=0.83\columnwidth]{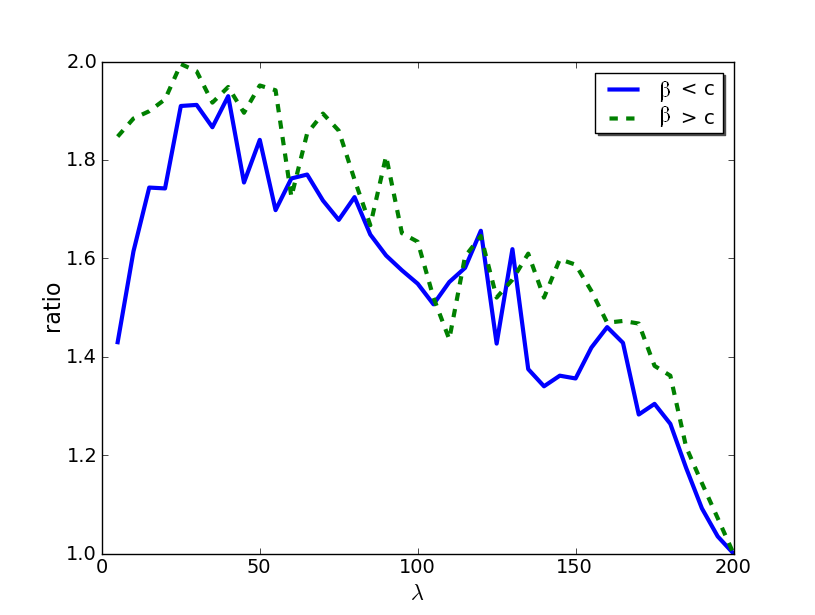}\\
\caption{Ratio between $\STAT$ and $\OPT$ cost as a function of
$\lambda$ in time zone scenario ($p=50\%$). Runtime 200 rounds,
three requests per round, network size five, and averaged over ten
runs.}\label{fig:opt_by_stat_commuter_dl_vary_frequency_combined}
\end{center}
\end{figure}

The dependency of the ratio between $\STAT$ and $\OPT$ costs on
$\lambda$ is more accentuated in the time zone scenario.
Figure~\ref{fig:opt_by_stat_commuter_dl_vary_frequency_combined}
shows that while a very high dynamic yields a moderate ratio, the
ratio goes up quickly already for small $\lambda$, and then the use
of migration declines more or less linearly with lower dynamics.
Interestingly, the two variants $\beta<c$ and $\beta>c$ yield
similar results in this time zone scenario. These results can be
explained as follows: for the commuter scenarios, new servers have
to be created rapidly as the requests fan out; migration does not
help much, and the load on the servers plays a minor role. Thus, for
$c < \beta$ we observe a better ratio between $\STAT$ and $\OPT$
(smaller cost $c$). For time zone on the other hand, the requests
move highly correlated, and there is not much of a difference
whether new servers are created or existing servers are migrated.
This explains why the ratio for $\beta<c$ is similar to the one for
$\beta>c$.

\begin{figure}[t]
\begin{center}
\includegraphics[width=0.83\columnwidth]{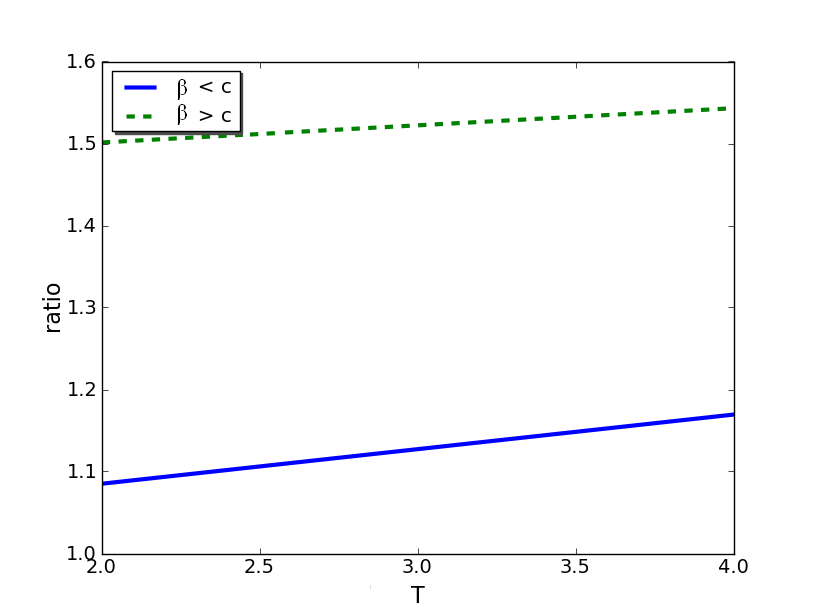}\\
\caption{Commuter scenario with dynamic load, where ratio of $\STAT$
and $\OPT$ costs are plotted as a function of $T$. Runtime 200
rounds, $\lambda=10$, network size five, averaged over ten
runs.}\label{fig:opt_by_stat_commuter_sl_vary_frequency_combined}
\end{center}
\end{figure}

\begin{figure}[t]
\begin{center}
\includegraphics[width=0.83\columnwidth]{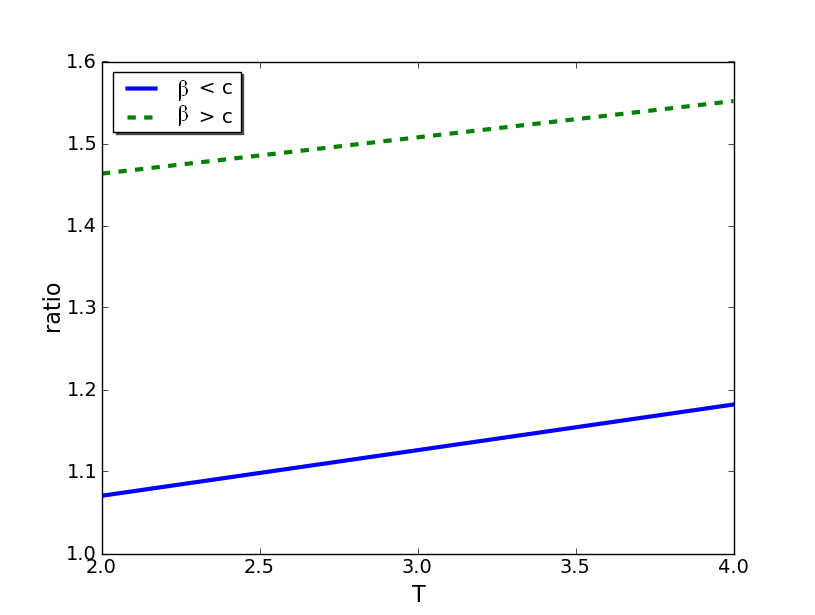}\\
\caption{As
Figure~\ref{fig:opt_by_stat_commuter_sl_vary_frequency_combined} but
for static load.}\label{fig:opt_vs_stat_commuter_dl_vary_frequency}
\end{center}
\end{figure}

We also conducted experiments studying the impact of $T$. Recall
that for larger $T$, the request horizon becomes larger, and hence,
we expect higher absolute costs as well as a higher benefit from
migration.
Figure~\ref{fig:opt_by_stat_commuter_sl_vary_frequency_combined}
shows the ratio for a dynamic load commuter scenario, and
Figure~\ref{fig:opt_vs_stat_commuter_dl_vary_frequency} shows the
corresponding results in a static load scenario. The two experiments
confirm our expectations, and also indicate that the $\beta>c$
variants typically benefit more from migration.




Finally, we briefly report on the results we obtained in the
Rocketfuel network AS-7018 of \texttt{ATT} under the time zone
scenario. ($c=400$, $\beta=40$ $R_a=2.5$, $R_i=0.5$, runtime 600
rounds, $\lambda=20$, $p=50\%$): the total cost of $\STAT$ was
$26063.8129053$. $\ONTH$ was a factor less than two higher (cost
44176.288923) while $\ONBR$ had costs $111470.296256$.

\section{Related Work} \label{sec_relwork}

Our work is related to past and ongoing research in several fields.
In the following, the literature in these fields is reviewed and
discussed individually.

\textbf{Network Virtualization:} It is expected that in the future,
virtual networks will be allocated, maintained and managed like
clouds, offering flexibility and elasticity of resources allocated
for a limited time and driven by the demand; indeed, the algorithms
in this paper are of applicable in cloud environments. Network
virtualization has gained attention~\cite{geni} because it enables
the co-existence of innovation and reliability~\cite{visa09virtu}
and promises to overcome the ``ossification'' of the
Internet~\cite{ossification}. For a more detailed survey on the
subject, please refer to~\cite{virsurvey}. Virtualization allows to
support a variety of network architectures and services over a
shared substrate, that is, a \emph{Substrate Network Provider (SNP)}
provides a common substrate supporting a number of \emph{Diversified
Virtual Networks (DVN)}. OpenFlow~\cite{openflow} and
VINI~\cite{Bavier06} are two examples that allow researchers to
(simultaneously) evaluate protocols in a controllable and realistic
environment. Trellis~\cite{trellis08} provides such a software
platform for hosting multiple virtual networks on shared commodity
hardware and can be used for VINI. Network virtualization is also
useful in data center architectures, see, e.g., \cite{secondnet}.

\textbf{Embedding:} One major challenge in this context is the
\emph{embedding}~\cite{embedding} of VNets, that is, the question of
how to efficiently and on-demand assign incoming service requests
onto the topology. Due to its relevance, the embedding problem has
been intensively studied in various settings, e.g., for an offline
version of the embedding problem see~\cite{turner}, for an embedding
with only bandwidth constraints see~\cite{ammar}, for heuristic
approaches without admission control see~\cite{zhu06}, or for a
simulated annealing approach see~\cite{simannealing}. Since the
general embedding problem is computationally hard, Yu et
al.~\cite{rethinking} advocate to rethink the design of the
substrate network to simplify the embedding; for instance, they
allow to split a virtual link over multiple paths and perform
periodic path migrations. Lischka and Karl~\cite{pb-embed} present
an embedding heuristic that uses backtracking and aims at embedding
nodes and links concurrently for improved resource utilization. Such
a concurrent mapping approach is also proposed in~\cite{infocom2009}
with the help of a mixed integer program. Finally, several
challenges of embeddings in wireless networks have been identified
by Park and Kim~\cite{wembed}.

\textbf{Migration and Allocation:} In contrast to the approaches
discussed above we, in this paper, tackle the question of how to
dynamically embed or migrate virtual servers~\cite{mobitopolo} in
order to efficiently satisfy connection requests arriving online at
any of the network entry points, and thus use virtualization
technology to improve the quality of service for mobile users. The
relevance of this subproblem of the general embedding problem is
underlined by Hao et al.~\cite{visa09mig} who show that under
certain circumstances, migration of a Samba front-end server closer
to the users can be beneficial even for bulk-data applications. Our
paper builds upon the single server architecture studied
in~\cite{visa10}, where a competitive online algorithm has been
described to migrate a single server depending on the dynamics of
mobile users. In contrast to~\cite{visa10} which attends to the
question of \emph{where} to migrate a server, we, in this paper,
initiate the study of when to allocate \emph{additional servers};
moreover, we generalize the migration model of~\cite{visa10} by
taking into account the running costs of a server. Also, in contrast
to~\cite{visa10} our model is aware of the load induced by the
traffic arriving at a node, and we present solutions for very
general load functions; for example, this implies that for scenarios
where the cost highly depends on the number of requests, relatively
more servers will be allocated to balance the cost. Note that how to
dynamically allocate and expand/release resources has been studied
in different contexts before as well. For instance,
\cite{visa10adaptive} describe a distributed virtual resource
provisioning and embedding algorithm based on an autonomous agent
framework, with a focus on fault-tolerance.

\textbf{Online Algorithms:} Due to the dynamic nature of virtual
networks, the field of online algorithms and competitive analysis
offers many tools that are useful to design strategies with
performance guarantees under uncertainty of the request pattern. For
instance, in the field of \emph{facility location}, researchers aim
at computing optimal facility locations that minimize building costs
and access costs (see, e.g.,~\cite{competitivefl} for an online
algorithm). As in our model, the uncapacitated online facility
location problem allows to create (but not shut down again!)
facilities when needed, and hence the set of ``resources'' is
dynamic. In contrast to the classic facility location problems, in
our model an additional server does not only come at a certain
creation cost, but also entails running costs; moreover, our model
incorporates a notion of mobility of requests, and servers can be
migrated and shut down again. In \cite{georgiosmigration}, a
heuristic algorithm is proposed for a variant of a facility location
problem which allows for facility migration; this algorithm uses
neighborhood-limited topology and demand information to compute
optimal facility locations in a distributed manner. In contrast to
our work, the setting is different and migration cost is measured in
terms of hop count. In the field of \emph{$k$-server problems}
(e.g.,~\cite{borodin}), an online algorithm must control the
movement of a set of $k$ servers, represented as points in a metric
space, and handle requests that are also in the form of points in
the space. As each request arrives, the algorithm must determine
which server to move to the requested point. The goal of the
algorithm is to reduce the total distance that all servers traverse.
In contrast, in our model the server can be accessed remotely, that
is, there is no need for the server to move to the request's
position. The \emph{page migration problem} (e.g.,
\cite{dynpmsurvey}) occurs in managing a globally addressed shared
memory in a multiprocessor system. Each physical page of memory is
located at a given processor, and memory references to that page by
other processors are charged a cost equal to the network distance.
At times, the page may migrate between processors, at a cost equal
to the distance times a page size factor. The problem is to schedule
movements on-line so as to minimize the total cost of memory
references. In contrast to these page migration models, we
differentiate between access costs that are determined by latency
and migration costs that are determined by network bandwidth. Most
of the models discussed are instances of so-called \emph{metrical
task systems}~\cite{borodin,metricaltask}) for which there is, e.g.,
an asymptotically optimal deterministic $\Theta(n)$-competitive
algorithm, where $n$ is the state space; or a randomized $O(\log^2
n\cdot \log\log{n})$-competitive algorithm given that the state
space fulfills the triangle inequality: this algorithm uses a (well
separated) tree approximation for the general metric space (in a
preprocessing step) and subsequently solves the problem on this
distorted space; unfortunately, both algorithmic parts are rather
complex. As pointed out in~\cite{visa10}, there is an intriguing
relationship between server migration and \emph{online function
tracking}~\cite{swat10,soda09}. In online function tracking, an
entity Alice needs to keep an entity Bob (approximately) informed
about a dynamically changing function, without sending too many
updates. The online function tracking problem can be transformed
into a chain network where the function values are represented by
the nodes on the chain, and a sequence of value changes corresponds
to a request pattern on the chain.

\end{document}